\documentstyle[12pt]{article}
\newcommand{\beq}{\begin{equation}}
\newcommand{\eeq}{\end{equation}}

\newcommand{\dd}{D\hspace{-.65em}/}

\def\op{operator}
\def\dop{Dirac operator}

\def\cn{condition}

\begin{document}
\begin{titlepage}
\begin{flushright}
NBI-HE-93-75 \\
December 1993\\
\end{flushright}
\vspace{0.5cm}
\begin{center}
{\large {\bf On sphaleron deformations
induced by Yukawa interactions}}\\
\vspace{1.5cm}
{\bf Minos Axenides}
\footnote{e-mail:axenides@nbivax.nbi.dk}\\
\vspace{0.4cm}
{\em The Niels Bohr Institute\\
University of Copenhagen, 17 Blegdamsvej, 2100 Copenhagen, Denmark}\\
\vspace{0.4cm}
{\bf Andrei Johansen}
\footnote{e-mail:johansen@lnpi.spb.su}\\
\vspace{0.4cm}
{\em The St.Petersburg Nuclear Physics Institute\\
 Gatchina, St.Petersburg District, 188350 Russia}\\
\vspace{0.4cm}
{\bf Holger Bech Nielsen}
\footnote{e-mail:hbech@nbivax.nbi.dk}\\
\vspace{0.4cm}
{\em The Niels Bohr Institute\\
University of Copenhagen, 17 Blegdamsvej, 2100 Copenhagen, Denmark}\\

\end{center}
\begin{abstract}
Due to the presence of the chiral anomaly
sphalerons with Chern-Simons number a half (CS=1/2) are
the only static configurations that allow for a fermion level
crossing in the two-dimensional Abelian-Higgs model with
massless fermions, i.e. in the absence of Yukawa interactions.
In the presence of fermion-Higgs interactions
we demonstrate the existence of zero energy solutions to the
one-dimensional Dirac equation at deformed sphalerons
with CS$\neq 1/2 .$
Induced level crossing due to Yukawa interactions
illustrates a non-trivial generalization of the Atiyah-Patodi-Singer
index theorem and of the
equivalence between parity anomaly in odd and
the chiral anomaly in even dimensions.
We discuss a subtle manifestation of this effect in the
standard electroweak theory at finite temperatures.

\end{abstract}
\end{titlepage}
\newpage
\section{Introduction}
\setcounter{equation}{0}

The standard model of electroweak interactions violates
baryon number through the chiral anomaly \cite{hooft}.
One of the reasons for this is the periodic structure of the vacuum
in non-abelian gauge theories.
The distinct ground states are labeled by integer
values of $N_{CS} ,$ the Chern-Simons number of the $SU(2)$
gauge field $A_{\mu}$
\beq
N_{CS} = \frac{g^2}{16\pi^2} \int_{M^3}
{\rm Tr} (AdA + \frac{2}{3} A^3)
\eeq
where $M^3$ is a 3-dimensional manifold.

The height of the potential barrier between adjacent vacua is given by
the energy of the sphaleron $E_{sp} = O(M_W/\alpha)$
($\alpha$ is $SU(2)$ gauge coupling).
It is a static and unstable solution to the classical
equations of the $SU(2)$ theory.
Any transition between vacuum sectors is accompanied
by a net change in both baryon and lepton
numbers due to level
crossing
\beq
\Delta B = \Delta L = n_f \Delta N_{CS} =
\frac{n_f}{32\pi^2} \int_{M^4} {\rm Tr} F^2 =
n_f \int_{M^4} \partial_{\mu} j_5^{\mu} \propto n_f (n_+ -n_-) ,
\eeq
where $M^4 = M^3 \times S^1$ is a four dimensional euclidean
manifold.
Here $n_f$ is the number of families,
$F = dA + A^2$ is the $SU(2)$ field strength and
$n_{\pm}$ is the number of left(right)-handed zero modes
of the 4-dimensional Dirac \op \ $i\dd_4 (A) .$
Such a transition is signaled by the presence of a zero energy
normalizable solution to the 4-dimensional Dirac equation (quantum
tunneling) or the 3-dimensional one in a sphaleron-like background
(thermal jump).

The anomaly of the chiral current becomes
an anomaly in the baryon and lepton currents because the standard model
is chiral.
At zero temperature and energy B-violating transitions
are exponentially suppressed.
At high temperature and in thermal equilibrium the
probability of finding a sphaleron-like configuration in the hot plasma
is given by the Boltzman weight of the sphaleron
$\exp (-E_{sp}/T) .$
At temperatures high compared to $M_W$ but low compared to $M_W/\alpha$
transitions over the barrier are
governed by classical statistical mechanics and become quite rapid
at $T \sim M_W/\alpha$ \cite{cohen}.
In this regime one is supposed to study time independent
solution to the equation of motion.
Three dimensional static gauge and Higgs fields of the sphaleron
type \cite{klink} and its various deformed versions \cite{yaffe}.
It should be noted however that fermion level crossing is not to be
restricted only to solutions to the electroweak equations of motion as
backgrounds to the Dirac equation.
The term ``sphaleron deformation'' will henceforth define the general
class of gauge and Higgs field sphaleron-like configurations that
admit level crossing.

Eq.(1.2) provides us with the sufficient condition for level crossing
to occur in the hot primordial plasma and in the presence
of large thermal fluctuations of the gauge and Higgs fields.
More precisely by adopting a ``real time evolution''
model of a thermal fluctuation of a classical
field we may consider a continuous string of static fields
$\phi (x,t),$ $A_i(x,t),$ $t\in [0,1]$ that interpolates between two
adjacent vacua with $t=0$ and $t=1 .$
(We may identify time with the Chern-Simons number and concentrate
on the $n=0,1$ sectors of the vacuum).
Provided that our set of static
configurations satisfies (in the $A_0 =0$ gauge)
appropriate boundary \cn s there will be a single level
crossing from the lowest positive fermion eigenstate
into the highest negative one.
These are given by
\beq
A_i =0,\;\;\; \phi =\phi_0 \;\;\; {\rm at} \;\; t=0 ,
\eeq
$$A_i = g^{-1} \partial_i g ,\;\;\;
\phi = g\phi_0 \;\;\; {\rm at} \;\; t=1 .$$
Here $g$ is a large gauge transformation and $\phi_0$ is a constant
$SU(2)$ doublet.

But at exactly which configuration and with what CS number
will this occur?
The issue from the phenomenological point of view may sound
inconsequential and academic.
Indeed in the minimal electroweak theory rapid fermion
level crossing is the one \cn \ for electroweak
baryogenesis in the early Universe believed
to be best understood and established.
We feel nevertheless that the elucidation of subtle aspects
of a well grounded effect to be always worthwhile.
In this spirit we will try to answer the previously posed question.
The topological \cn \ (1.2) does not single out any specific
sphaleron-like configuration.
Indeed for the case of massless fermions and in the absence
of Yukawa interactions numerical studies have corroborated to
this conclusion \cite{ambj} while explicit
fermion zero mode constructions have been given only for
in the specific sphaleron
backgrounds with CS$=1/2$ \cite{kunz}.

The case of Yukawa interactions however appears to be somewhat
more subtle.
The presence of a zero in the Higgs field and the fermion
mass matrix appears to be an additional necessary \cn \
intuitively for level crossing to occur.
Indeed this appeared to be the case at least for the sphaleron case
in a recent study \cite{mah}.
Moreover one can observe in the same treatment of the problem
that in the presence of a constant Higgs field the normalizability
of the zero mode is lost.
By analogy with the massless fermion case discussed previously
it would be reasonable to expect this physical picture to hold true
for a sphaleron like configuration with CS$\neq 1/2 .$
Surprisingly recent numerical simulations \cite{amb}
reveal that there exist sphaleron deformations with CS$\neq 1/2$
in the presence of Yukawa interactions and a constant Higgs field
throughout space, which allow for level crossing at least for small
values of Yukawa couplings.
In the present note we take up again the investigation
of the effect of Yukawa interactions on the level crossing in
the two dimensional Abelian-Higgs model.

We do it by introducing the Atiyah-Patodi-Singer (APS) index
(the $\eta$ invariant) that measures the spectral asymmetry
of a 3-dimensional Dirac operator in a continuous set of external
static gauge + Higgs fields $A_t ,$ $\phi_t$ with $t\in [t_0 ,t_1]$
\cite{aps}.
It is defined to be
\beq
\eta = \sum_{\lambda_k > 0} 1 -
\sum_{\lambda_k < 0} 1 =
\eeq
$$= \sum_k \; ' \frac{{\rm sign}\; \lambda_k}{|\lambda_k|^s} \;\;\;
{\rm at} \;\;\; {\rm Re} \; s>0, \;\;\; s\to 0,$$
where the second line gives a
regularized expression for the spectral asymmetry.
The symbol $\sum'$ implies a sum over all
non-zero eigenvalues $\lambda_k$ of the \dop \ and
$s$ is taken to be a complex parameter.
Here the boundaries of the set of fields are not
to be identified with the vacuum states with CS$=0$ and 1 respectively.
More geometrically they are the three dimensional connections on a
manifold of a cylinder $M^3\times I$ with $t$ parametrizing
$I=[t_0 ,t_1] .$
The APS index theorem then \cite{aps} implies
that if the eigenfunctions of a properly defined 4-dimensional
$\dd_4 = \partial/\partial_t +\dd_3$ \dop \
are to be normalizable on $M^3\times I$ then
\beq
\eta(A_{t_1}) - \eta(A_{t_0}) = \frac{1}{16\pi^2}
\int_{M^3\times I} {\rm Tr} F^2 -2q ,
\eeq
where $q$ stands for Pontryagin index here taken to be 1.
It can be shown that when an eigenvalue of $\dd_3 (A_t)$
crosses the zero axis at some $t=t^*$ the spectral
asymmetry varies discontinuously by 2 units.
It thus has a continuous ($\eta_c$) and discontinuous parts
(spectral flow $[t^*]$).
The latter is twice the number of zero level crossings
(see Fig.1)
\beq
\eta(A_{t_1}) - \eta(A_{t_0}) = \int^{t_1}_{t_0} dt \frac{d}{dt}
\eta_c [A_t] + 2 \times {\rm spectral}\;\; {\rm flow} \; [t^*].
\eeq
It can further be shown that
\beq
\int^{t_1}_{t_0} dt \frac{d}{dt} \eta_c [A_t] =
\frac{1}{16\pi^2} \int_{M^3 \times I} {\rm Tr} F^2 .
\eeq
We may observe two interesting limits.
In the case where
the boundaries $A_0 ,$ $A_1$ are identical to the vacua at CS$=0$ and 1
respectively we recover the usual 4-dimensional Pontryagin index
relation.
In the opposite limit where $t_0\sim t_1 \sim t^*$ we have that
$$-{\rm spectral \;\; flow} \; = \;
{\rm topological \;\; charge}\;\; q=1 .$$
Interestingly both the 3-dimensional spectral flow and the
4-dimensional index probe the same homotopic group
$\pi_3 (SU(2)) = {\bf Z} .$
The spectral flow across the static configuration $A_{t^*}$
can be computed by its parity anomaly \cite{moore,forte}.
In equations $\eta \sim {\rm Im} W(A).$
The $\eta$ invariant is proportional to the parity anomaly
which is given by
the imaginary part of the effective action
of the 3-dimensional configuration of the
gauge field.

In section 2 we illustrate the general
formalism by applying it to the two dimensional
Abelian-Higgs model with massless fermions and no
Yukawa interactions.
We demonstrate spectral flow of the $\eta$ invariant at sphaleron
configurations of CS =1/2.
In section 2 we investigate the effect of Yukawa interactions.
We find that a constant Higgs field with no zero is compatible
with the existence of a normalizable zero mode of the one dimensional
\dop .
Interestingly this contrasts with the electroweak sphaleron case
in 3+1 dimensions about which we made some remarks before.
Furthermore
Yukawa interactions may induce sphaleron deformations with CS$\neq 1/2$
which possess normalizable zero modes
and hence allow for level crossing as well.
We finally examine the presence of both effects in
the realistic electroweak theory.

\section{Abelian-Higgs model with massless fermions}
\setcounter{equation}{0}

Here we consider an Abelian-Higgs model as
a simple example which demonstrates the role of parity
anomaly for the description of the level crossing phenomenon.

We consider the euclidean version of the theory
whose Lagrangian reads as follows \cite{boch}
\beq
L = \psi^+_R i\bar{D}\psi_L + \psi^+_L iD \psi_R .
\eeq
Here $D = \partial -iA ,$
 $\bar{D} = \bar{\partial}
-i\bar{A},$ $\partial = \partial /\partial z ,$
$\bar{\partial} = \partial /\partial \bar{z},$ and $z=x_1+ix_2,$
$\bar{z} = x_1 -ix_2,$ $A=A_1 -iA_2,$ $\bar{A} = A_1 +iA_2 .$
The fermionic fields $\psi_L,\; \psi_R,\; \psi^+_L,$
and $\psi^+_R$ are chiral components of the
corresponding Dirac spinors.
For definiteness
we fix that $x_1$ is ``time'', while $x_2 =x$ is a (space)
coordinate on a circle.

The $\eta$ invariant is a functional of a 0+1 dimensional
external gauge field.
To calculate the $\eta$ invariant let us
choose the Hamiltonian gauge ($A_1 =0$) and write down the
0+1 dimensional Dirac operator for fermions with a definite chirality
\beq
D_x = \partial_x -iA_x.
\eeq
It is easy to find the spectrum of this operator.
The equation for its eigenvalues $\lambda$ is given by
\beq
(\partial_x -iA_x) \psi_{\lambda} = i \lambda \psi_{\lambda} .
\eeq
We fix the anti-periodic
boundary \cn \ for the corresponding wave function on a circle.
It reads as
\beq
\psi_{\lambda}(x) = \exp (\int^x_{-L/2} A_x dx + i\lambda x)
\psi_0 ,
\eeq
where $\psi_0$ is a non-vanishing constant.
Using the anti-periodicity \cn \ we get
\beq
\lambda_n = -2\pi a/L
+ \pi/L + 2\pi n/L,
\eeq
where $n \in {\bf Z} ,$
and $a =\frac{1}{2\pi} \int^{L/2}_{-L/2} dx \; A_x $ is the Chern-Simons
functional for the $(0+1)$ case.

For definiteness we assume that
$0<a<1/2$ so that we restrict ourselves to computing the continuous
part of $\eta .$
The partition function now reads \cite{jackiw}
\beq
Z= const \times
\prod^{\infty}_{n= -\infty} (n +\frac{1}{2} - a)=
\cos \; a\pi .
\eeq
Here we introduced a constant which does not depend on the external
gauge field.
It is clear that this result is not gauge invariant as it changes its
sign when one makes a big gauge transformation $a\to a + 1 .$
This is actully a result of a special choice of regularization as
discussed in ref.\cite{schwimmer}.
With an appropriate regularization
we can still restore the gauge invariance.
In this case we effectively get an additional counter term proportional
to the Chern-Simons functional $a .$
Simultaneously we loose the parity invariance of the theory
as can be seen from the modified partition function
\beq
Z= \cos\pi a \; e^{ia\pi} .
\eeq
Now we can extract the imaginary part of this partition function
to be identified with a parity anomaly.
It is given by Im $\log Z$ as follows
\beq
\eta[A]= \frac{2}{\pi} {\rm Im} \log Z = 2a + \eta_{step} ,
\eeq
where $\eta_{step}$ is a discontinuous part of the $
\eta$ invariant.
This expression is gauge invariant under big gauge transformations
due to the discontinuous part $\eta_{step} .$
It is easy to see that the jumps of $\eta_{step}$ correspond
to half integer values of the Chern-Simons
functional $a =n +1/2,$ $n\in {\bf Z} .$

In what follows we present a different but equivalent way
to compute the parity anomaly through the spectral asymmetry
of the \dop .
Using the $\zeta$ regularization we have \cite{moore}
\beq
\eta[A] = \sum_n
\frac{{\rm sign} \lambda_n}{|\lambda_n|^{s}}|_{s\to 0} .
\eeq
Taking once again for definiteness $0<a<1/2$
we get
\beq
\eta[A] = 1 + \sum^{\infty}_{n=1} \left( \frac{1}{|n-a+1/2|^{s}} -
\frac{1}{|n+a-1/2|^{s}} \right) .
\eeq
To determine the dependence of this functional on $a$
we calculate its derivative
\beq
\frac{\partial}{\partial a} \eta [A] =
\sum_{n=1}^{\infty} \left( \frac{s}{(n-a+1/2)^{s+1}}+
\frac{s}{(n+a-1/2)^{s+1}} \right) .
\eeq
This infinite sum diverges at $s\to 0$ and to the leading order
it is given by $1/s .$
We thus get the same result for the continuous part of the
parity anomaly as above.
The discontinuous part obviously appears in such a calculation when
we accurately take into account first terms in the sum in eq.(2.9).

Finally let us calculate the phase shift in the APS theorem
for this model.
Indeed we may use the Ward identity for the $\gamma_5$ current.
We define the $\gamma_5$ current as follows
\beq
j_5^{\mu} = \psi^+ \gamma_{\mu}\gamma_5 \psi =
\psi^+_R \gamma_{\mu}\psi_L -
\psi^+_L \gamma_{\mu}\psi_R ,
\eeq
where $\gamma_1 = \sigma_1$ and $\gamma_2 = \sigma_2 ,$ while
$\gamma_5 =\sigma_3 .$
This current has an anomaly and
the anomalous Ward identity for its divergence reads
\beq
<\partial_{\mu} j_{\mu} = \frac{i}{2\pi} \epsilon^{\mu\nu}
F_{\mu\nu} + 2m\psi^+ \gamma_5 \psi > .
\eeq
Integrating over $x$ and $t$ the last term in the right hand side
gives the difference of left- and right- handed fermionic zero modes.
Usually by assuming that the external field rapidly decreases at
infinity one can easily deduce from the above equation
the Atiyah-Singer
index theorem.
In that case the two dimensional integral
of the divergence of the axial current should be zero.

If instead we assume that the external field does not vanish
this term makes a non-zero contribution to the left hand side.
Let us assume that the external field rapidly decrease
at large $x ,$ while it does not vanish as $t \to \infty .$
In this case we get
\beq
\frac{1}{i}
<Q_5 (t=-\infty) -Q_5 (t=+\infty)>= \int dx \frac{1}{2\pi}
\epsilon^{\mu\nu} F_{\mu\nu} + 2(n_+ - n_-) ,
\eeq
where $n_+$ ($n_-$) are the numbers of left (right)-handed
fermionic zero modes with the chiral charge $Q_5$ being
defined as follows
\beq
Q_5 = \int dx \psi^+ \gamma_1 \gamma_5 \psi .
\eeq
Let us now calculate the
expectation value of $Q_5$ in the presence of a gauge field
\beq
<Q_5> =\int dx <\psi^+ \gamma_1 \gamma_5 \psi> =
\eeq
$$=\int dx <{\rm Tr}
(G_{RL}(x,y)e^{-i\int^x_y dx'_{\mu} A_{\mu}}
-G_{LR}(x,y) e^{-i\int^x_y dx'_{\mu} A_{\mu}})>|_{x\to y} .$$
Here the Green functions $G_{RL}(x,y)$ and $G_{LR}(x,y)$
obey the equations
\beq
i\bar{D} G_{RL}(x,y) = \delta^2 (x-y) ,
\;\;\; iD G_{LR}(x,y) = \delta^2 (x-y).
\eeq
The exponentials above are introduced
in order maintain the gauge invariance
under regularization (point splitting).
It is convenient to use the following representation for the gauge field
\beq
A_{\mu} = \partial_{\mu} \alpha + \epsilon_{\mu\nu} \partial^{\nu}
\beta .
\eeq
Here $\alpha$ and $\beta$ are scalar functions.
Then it is easy to check that
\beq
G_{RL}(x,y) =\frac{1}{2i\pi}
e^{i(\bar{\phi}(x) - \bar{\phi} (y))} \frac{1}{\bar{z}-\bar{z}'}
,\;\;\;
G_{LR}(x,y) = \frac{1}{2i\pi}
e^{i(\phi (x) - \phi (y))} \frac{1}{z- z'},
\eeq
where
\beq
\bar{\phi} = \alpha + i\beta ,\;\;\; \phi = \alpha - i\beta ,
\eeq
and $z= x_1 +ix_2 ,$ $ z'= y_1 +iy_2 .$

Taking the limit $x\to y$ we get
\beq
<Q_5> = - \frac{i}{\pi} \int dx \partial_1 \beta
= -2ia=-i\eta[A].
\eeq
This expression is proportional to the Chern-Simons functional $a=
\frac{1}{2\pi} \int dx A_2
= \frac{1}{2\pi}
\int dx_2 (\partial_2 \alpha - \partial_1 \beta)$ for periodic $\alpha .$
We thus get once more the same result for the $\eta$ invariant.

\section{Abelian-Higgs model with Yukawa interactions}
\setcounter{equation}{0}

Let us now consider the level crossing phenomenon in the presence of
Yukawa couplings.

The lagrangian now reads
\beq
L = \psi^+_R i\bar{D} \psi_L + \eta^+_L i \partial \eta_R +
i H \psi^+_R \eta_R + iH^* \eta^+_L \psi_L .
\eeq
Here $H$ is a complex Higgs field which can have a non-vanishing
vacuum expectation value at infinity.
In this case the $U(1)$ gauge invariance is spontaneously broken.
An additional neutral fermion is introduced to generate a mass
of the fermion $\psi .$
The lagrangian above is of course anomalous at the quantum level
and therefore we shall assume that there is another charged fermion
with the same charge but of an opposite chirality.

We take the same sphaleron configuration as above and the Hamiltonian
gauge.
The reduced $1+0$ lagrangian now reads
\beq
L = -\psi^+_R D_x \psi_L + \eta^+_L \partial_x \eta_R +
i H \psi^+_R \eta_R + iH^* \eta^+_L \psi_L .
\eeq
We are unable to find the spectrum for this particular
Hamiltonian.
We observe, however, that for a given value of the $\eta$-invariant
there exist a continuous set $\Omega$ of Hamiltonians that gives
rise to it.
Each member is related to the other by a constant matrix.
On the basis of this property of the $\eta$-invariant we pick
a more manageable Hamiltonian for which we can find its spectrum
and determine our sought after $\eta$-invariant.
Specifically our $\eta$-invariant is given by the imaginary part
of the logarithm of the fermionic determinant
($\eta \propto {\rm Im} \det \dd (A)$).
We observe that the value of $\eta$ is the same if we choose
$\dd \; '(A) = \dd (A) \tau_3 ,$ where $\tau_3$ is a Pauli matrix.
In our particular model (eq.(3.2)) we compute the $\eta$-invariant
by redefining our \dop \ in the way indicated above.
This is accomplished by effectively changing the sign of $\psi_L .$
Our modified Dirac equation is given by
\beq
D_x \psi_L + iH \eta_R = i\lambda \psi_L ,
\eeq
$$\partial_x \eta_R - iH^* \psi_L = i\lambda \eta_R .$$
The Dirac \op \ has the form
\beq
\hat{D} = \left( \begin{array}{cc} D_x & iH \\
- iH^* & \partial_x \end{array} \right) .
\eeq
This \op \ is not anti-hermitean, but
its eigenvalues will be shown to be purely imaginary.

It is now convenient to remove the gauge field from the
covariant derivative by a gauge rotation of the wave
function
\beq
\psi_L =g(x)\psi(x) =
e^{i\int^x_{L/2} dx \; A_x} \psi ,\;\;\; \eta_R = \eta .
\eeq
Now it easy to see that the (rotated) Higgs field enters as
external non-abelian gauge field in the equation.
However this "gauge" field is anti-hermitean.
Thus the general solution to the Dirac equations
reads
\beq
\psi_L =  e^{i\lambda x} \left( \begin{array}{cc} g(x) & 0\\ 0& 1
\end{array} \right)
U(x) q_0 ,
\eeq
where $q_0$ is a constant two-components wave function for $\psi_L$ and
$\eta_R$ fermions,
and
\beq
U(x) = P \; \exp (-\int^x_{-L/2} B) ,
\eeq
$$B=\left( \begin{array}{cc} 0& iHg^*\\
-iH^* g & 0 \end{array} \right) .$$
The matrix $B$ is hermitean, and $P$ stands for path ordering.

In order to determine the eigenvalues
we should take into account the anti-periodic
boundary \cn s for fermions
\beq
e^{-i\lambda L/2} q_0 =
e^{i\lambda L/2} \left( \begin{array}{cc} g(L/2) & 0\\ 0& 1
\end{array} \right) U(L/2) q_0 ,
\eeq
It is easy to see that
\beq
\det U(L/2) =1 ,
\eeq
since the matrix $B$ is traceless.
Moreover the diagonal matrix elements of the matrix
\beq
U(L/2) = \left( \begin{array}{cc} x & y \\ z & v \end{array} \right)
\eeq
are complex conjugated
\beq
x= r e^{i\theta} , \;\;\; v=r e^{-i\theta} ,
\eeq
where $r$ and $\theta$ are real.
The equation for the eigenvalues takes the form
\beq
\det \left[ e^{i\lambda L} \left(\begin{array}{cc} g &0 \\0 & 1
\end{array} \right)
\left(\begin{array}{cc} x & y \\z & v
\end{array} \right) + 1 \right] =0 .
\eeq
After some calculation we get
\beq
e^{-2i\lambda L} + e^{-i\lambda L}(v + xg) + g =0 .
\eeq
In particular for a zero mode ($\lambda =0$) we have
\beq
1 + g + x^* + xg =0 .
\eeq
This equation fixes the value of the Chern-Simons number $a$ for which
the level crossing occurs
\beq
a = \frac{1}{2} + \frac{1}{2\pi} {\rm arctan}
\frac{r \sin \theta}{1+ r\cos \theta} .
\eeq
It is clear that the angle $\theta$ can have any value.
We call this phenomenon, shift of $a$ from 1/2, {\em induced} level
crossing.
In this sense level crossing is purely an effect of
the presence of Yukawa interactions.
This is indeed the case
as in their absence level crossing can only occur
for configurations with $a =1/2 .$
Clearly (3.15) implies that in their presence, it may occur
elsewhere, e.g. for $a=0 .$
For example a zero mode can very well exist at $a =0 .$

It is interesting now to make a direct calculation of the parity
anomaly in the presence of Yukawa interactions.
The quadratic equation above has two real solutions.
Therefore there are two branches of the spectrum
\beq
\lambda^{(+)}_n = \lambda^{(+)}_0 +2\pi n ,\;\;\;
\lambda^{(-)}_n = \lambda^{(-)}_0 +2\pi n ,
\eeq
where $n$ is integer, and
\beq
\lambda^{(\pm )}_0 = \frac{\pi a}{L} +\frac{\pi}{L} \pm {\rm arccos}
(r \cos (\theta -\pi a)).
\eeq
The eigenvalues $\lambda^{(\pm)}_n$ are real for $|r \cos (\theta -\pi
a)| \leq 1 .$
In this case an appropriate representation for the $\eta$-invariant
is given by
\beq
\eta = \sum_{n = -\infty}^{+ \infty}
\frac{{\rm sign} \lambda^{(+)}_n}{|\lambda^{(+)}_n|^s} +
\sum_{n = -\infty}^{+ \infty}
\frac{{\rm sign} \lambda^{(-)}_n}{|\lambda^{(-)}_n|^s} .
\eeq
As we are interested in the more general case where $\lambda_n^{(\pm)}$
are complex ($|r \cos (\theta -\pi a)| >1$) we identify the
$\eta$-invariant with the imaginary part of the logarithm of the
determinant (${\rm Im} \log \det \hat{D}$) of the \dop \ which is
given by eq.(3.4).
Its evaluation gives us a similar result with the massless case,
namely
\beq
\eta = \lambda^{(+)}_0 + \lambda^{(-)}_0 + {\rm const.}
=2a +{\rm const.}
\eeq
This a result agrees with our argument that the continuous part of
$\eta$ invariant is essentially proportional to
the Chern-Simons number and independent of the Yukawa couplings.

We illustrate level crossing induced by Yukawa
as it is depicted in eq.(3.15) by a simple example.
We take the Higgs field $H(x)$ to be a step
function (Fig.2) so that $H=-ib_1 \exp i2\pi a$
for $-L/2 <x<0$ and $H=-ib_2\exp i2\pi a$ for
$0<x<L/2$, $b_{1,2}\in{\bf C} .$
Then the matrix $U(L/2)$ takes the following form
\beq
U(L/2) = \exp \frac{L}{2} \left(
\begin{array}{cc} 0 & b_1 \\ b_1^+ & 0 \end{array} \right)\times
\exp \frac{L}{2} \left(
\begin{array}{cc} 0 & b_2 \\ b_2^+ & 0\end{array} \right) .
\eeq
{}From this equation we easily get
\beq
\theta = {\rm arg} (\cosh|b_1|\cosh|b_2|
-\sinh|b_1|\sinh|b_2| e^{i\gamma}) ,
\eeq
where
\beq
e^{i\gamma} = \frac{b_1 b^+_2}{|b_1||b_2|} .
\eeq
It is easy to check that changing the values of the parameters
$b_{1,2}$ and $\gamma$ the value of the Chern-Simons number
at which the level crossing happens can be arbitrarily shifted.

\section{Spectral flow across the electroweak sphaleron}
\setcounter{equation}{0}

We can now attempt to use the concept of spectral flow across
a static 3-dimensional configuration, that of the electroweak
sphaleron.
In the case of massless fermions we may again consider a
continuous set of static gauge + Higgs fields
$(\phi(x,t),\; A(x,t))$ parametrized by $t\in (0,1) .$

The APS index theorem (eq.(1.5)) is valid.
We notice that the continuous part of the $\eta$ invariant
is given by
\beq
\eta_c(A_t) \propto \int^t_0 dt\;\; {\rm Tr} F^2 .
\eeq
More precisely we take in eq.(1.7) $t_0 =0$ and $t_1 =t .$
Our experience with the 2-dimensional Abelian-Higgs model suggests
that if we choose an interval in $t$ that excludes a discontinuous
jump for the $\eta$ invariant, i.e. consider only the behaviour
of its continuous part $\eta_c (A_t)$ that should be proportional
to the Chern-Simons functional for $A_t$ plus a constant.
We now want to extend this argument to three dimensions.

We choose a gauge for the 3-dimensional gauge fields in a $t$-interval
that does not include any discontinuities for the $\eta$ invariant
so that they rapidly decrease at spatial infinity.
As before the $\eta_c (A_t)$ must be proportional to the
Chern-Simons functional for $A_t$ (plus a constant).
Furthermore a discontinuity in the $\eta$ inavariant across a particular
sphaleron like configuration $(\phi_{sp} (x, t^*), \;
A_{sp} (x,t^*))$ signals similarly a level crossing,
the existence of a normalizable zero energy solution, and hence
fermion level crossing.
As we previously stated this may occur at configurations with
CS$\neq 1/2 .$

It is less trivial to extend the same argument to the case of Yukawa
interactions with massive fermions.
Indeed in this case the $\eta$ invariant
in general depends on Yukawa couplings and Higgs fields.
Numerical simulations \cite{amb}
suggest surprisingly that for small Yukawa couplings there exist
deformed sphalerons that allow for level crossing without
the necessary presence of the zero of the Higgs field.
For larger values it was found, perhaps not so surprisingly,
that the situation is similar to the sphaleron case \cite{mah}.
Deformed sphalerons with CS$\neq 1/2$ admit a fermion zero mode
in the presence of a zero of Higgs field.
The Higgs field in three dimensions appears to have an important
and subtle bearing on the level crossing phenomenon.
This calls for a nontrivial generalization of our previous argument
for the APS index theorem, which we do not possess at the moment.
We may offer nevertheless some argument for the general behaviour
of the continuous part of the spectral asymmetry that we feel
holds true in this case too.
More specifically we argue that
it is independent of the Higgs configuration and
proportional to some Chern-Simons functional.
Indeed in the presence of Yukawa interactions a generalization
of the Atiyah-Singer index theorem has been demonstarted \cite{aj},
$2q = \frac{1}{16\pi^2} \int {\rm Tr} F^2 .$
We can now put it in the language of the APS index theorem
(eqs.(1.5),(1.7)).
Indeed from eq.(1.7) we can rewrite eq.(1.5) as follows
\beq
\frac{1}{16\pi^2} \int_{M^3 \times [1,t_1]} {\rm Tr} F^2 -
\int_{M^3 \times [t_0,0]} {\rm Tr} F^2 =
\int_{M^3 \times [t_0,t_1]} {\rm Tr} F^2 - 2q .
\eeq
Here too $t_0$ and $t_1$ are the boundaries of the continuous set
of gauge and Higgs fields $(A_i (x,t) , \; \phi(x,t)),$ $t\in [0,1] .$
Moreover $A_i (x,t_{0,1})$ and $\phi (x, t_{0,1})$are not the vacua with
CS =0,1 respectively.
It is obvious that from eq.(4.2)
\beq
\frac{1}{32\pi^2}\int_{M^3 \times [1, t_1]} {\rm Tr} F^2 = -1
+\frac{1}{32\pi^2}\int_{M^3 \times [0,t_1]} {\rm Tr} F^2 .
\eeq
A direct comparison with eq.(4.1) suggests that in fact eq.(4.2) is a
restatment of the APS index theorem with the $\eta_c [A_t]$ being
independent of the Higgs field and Yukawa couplings.
This is not to be taken of course as a demonstration that the
$\eta$-functional is given by the parity anomaly in the presence
of Yukawa interactions.
We believe this to be true here too.

We just stated there exist deformed
sphalerons with CS$\neq 1/2$ which possess zero energy
normalizable solution to the 3-dimensional Dirac equation.
In fact for small values of the Yukawa couplings the presence
of a zero in the Higgs field and fermion mass matrix is
unnecessary.
Surprisingly this is not the case with the sphaleron configuration
itself with CS=1/2
in contrast with the 2-dimensional Abelian-Higgs counterpart
configuration.
While this paradoxical difference is obscure to us we would nevertheless
like to
discuss the interesting possibility
that the Yukawa interactions in analogy with the 2-dimensional
model induce new sphaleron deformations that allow
for level crossing.
This certainly presupposes
a physical criterion (e.g. symmetry) that can classify and distinguish
the sphaleron deformations induced by Yukawa interactions
from the ones that exist in their absence and allow for level
crossing.
While we do not possess such an analytical tool at the moment
we point out that this is a formidable task too for
numerical simulations of the type that recently revealed
a shift in fermion level crossing due to Yukawa interactions.
To interprete such a shift as a sphaleron deformation induced
purely by them one must be able to exhaustively
trace all possible paths of sphaleron-like transition that
interpolate between two topologically distinct vacua
in the case of massless fermions and no Yukawa interactions
which moreover admit zero modes at deformed sphaleron backgrounds
with CS$\neq 1/2 .$

\section{Conclusions}
\setcounter{equation}{0}

In the present work we explored the effect of Yukawa interactions
and massive fermions on fermion level crossing
in the two dimensional Abelian-Higgs model.
In the massless fermion case sphalerons with CS$=1/2$ are
the only static configurations that allow for level crossing.
Consequently the existence of fermion zero modes
for deformed sphalerons with CS$\neq 1/2$ we interprete as
a clean signature of a sphaleron deformation induced by Yukawa
interactions.
In the realistic case of the electroweak sphaleron-like
transitions in the hot early Universe such an interpretation
is not straightforward
as in the massless fermion case level crossing can occur at deformed
sphaleron configuration with CS$\neq 1/2 .$
Our lack of any symmetry classification for such deformations
renders the interpretation of the observed shift
in level crossing \cite{amb} as giving rise to novel deformed
sphaleron configurations induced purely by Yukawa interactions
premature.

Our clean demonstration of such a possibility in 2 dimensions, however,
also points to a non-trivial generalization of the Atiyah-Patodi-Singer
(APS) index theorem along with the equivalence between the parity
anomaly of the sphaleron and the chiral anomaly for the case of Yukawa
interactions.

\section{Acknowledgments}

We are grateful to Jan Ambj\o rn for inspiration and suggestions.
We also thank him and Gudmar Thorleifsson
for communicating their numerical simulation results before
publication.
This research is partially supported by a NATO fund.
M.A. thanks the Carlsberg Foundation for financial support.

\newpage
\subsection*{Figure Captions}
\vskip 6pt

$\;\;\;\;\;$Fig.1. The spectral flow as denoted by the jumps of $\eta
(t) .$
Wherever a Dirac eigenvalue $\lambda$ crosses zero from
$\lambda > (<) 0$ levels to $\lambda < (>) 0$ levels the
$\eta$-invariant jumps by -(+) 2 units.

Fig.2. The APS index (continuous and discontinuous) for the
2-dimensional massless chiral Abelian-Higgs model.
At the sphaleron configuration with CS=1/2 occurs level crossing.

Fig.3. A trial Higgs field $H(x)$ step function configuration in
an 2-dimensional Abelian-Higgs model with Yukawa interactions.
Induced level crossing occurs at $x=0 .$

\end{document}